\documentclass[aps,preprint,groupedaddress,showpacs,floatfix]{revtex4}
\usepackage{graphicx}
\begin{document}

\title{Comparative study of the production of scalar and tensor mesons in $e^+e^-$ collisions}
\author {
N.N. Achasov$^{\,a}$ \email{achasov@math.nsc.ru}, A.I.
Goncharenko$^{\,a,b}$, A.V. Kiselev$^{\,a,b}$,
\email{kiselev@math.nsc.ru} and E.V. Rogozina$^{\,a,b}$}

\affiliation{
   $^a$Laboratory of Theoretical Physics,
 Sobolev Institute for Mathematics, 630090, Novosibirsk, Russia\\
$^b$Novosibirsk State University, 630090, Novosibirsk, Russia}

\date{\today}

\begin{abstract}

The intensity of scalar $a_0(980)$, $f_0(980)$ and tensor
$a_2(1320)$, $f_2(1270)$ mesons production at VEPP-2000 (BINP,
Novosibirsk) and the upgraded DA$\Phi$NE (Frascati, Italy) in the
processes $e^+e^- \to a_0 (f_0$, $a_2$, $f_2) \gamma$ is
calculated. The calculation was made with the help of VDM and the
kaon loop model (for scalar mesons case). Note that the processes
$e^+e^-\to a_2(f_2)\gamma$ have not been studied in the energy
region of VEPP-2000.

It turned out that in the VEPP-2000 energy region $1.7-2.0$ GeV
$\sigma(e^+e^-\to a_2(f_2)\gamma)\sim 10$ pb, and
$\sigma(e^+e^-\to a_0(f_0)\gamma)\sim 0.1$ pb.

Photon angle distribution and the spin density matrices of $a_2$
and $f_2$ were calculated also.

\end{abstract}
 \maketitle

\section{Introduction}

Study of the nature of light scalar resonances is one of the
central problems of nonperturbative QCD, it is important for
understanding the chiral symmetry realization way in the low
energy region and, consequently, the confinement physics. The
$a_2(1320)$ and $f_2(1270)$ tensor mesons are well-known P-wave
$q\bar q$ states. Naively one might think that the scalar
$a_0(980)$ and $f_0(980)$ mesons are also the $q\bar q$ P-wave
states with the same quark structure, as $a_2(1320)$ and
$f_2(1270)$, respectively. But now there are many indications that
the above scalars are the four quark states.

Comparative study of the production of scalar and tensor mesons is
proposed to investigate the nature of light scalar mesons. For
this purpose the intensity of scalar $a_0(980)$, $f_0(980)$ and
tensor $a_2(1320)$, $f_2(1270)$ mesons production at the colliders
VEPP-2000 (BINP, Novosibirsk) and DA$\Phi$NE (LNF, Frascati) in
the processes $e^+e^- \to S \gamma, T\gamma$ (here and hereafter
$S = a_0$, $f_0$; $T=a_2$, $f_2$) is calculated.


\section{The reactions $e^+e^-\to a_0\gamma \to \eta\pi^0\gamma$ and $e^+e^-\to f_0\gamma \to \pi^0\pi^0\gamma$}

As it is known, the kaon loop model \cite{achasov-89,achasov-97}
describe the $V\to K^+K^-\to a_0(f_0)\gamma$ decays well
\cite{snd,cmd,kloea0,kloef0,
a0f0,our_a0,achasov-03,our_f0,our_f0_2011}. The signal
contribution is $e^+e^-\to \sum_V V\to K^+K^-\to a_0(f_0)\gamma \to
\eta(\pi^0)\pi^0\gamma$, where $V$ are vector mesons. The signal
cross-sections

$$\frac{d\sigma (e^+e^-\to a_0\gamma\to
\eta\pi^0\gamma)}{dm}=\frac{g_{a_0 \eta\pi^0}^2
p_{\eta\pi}(m)}{4\pi^2|D_{a_0}(m)|^2}\,\sigma (e^+e^-\to
a_0\gamma,m)=$$
\begin{equation}
\frac{e^2 (s-m^2)p_{\eta\pi}(m)}{96\pi^3 s^3} |A_{K^+K^-}(s) \bar
g(m)|^2\bigg|\frac{g_{a_0 K^+K^-}g_{a_0
\eta\pi^0}}{D_{a_0}(m)}\bigg|^2 \,,\label{sigmaSignala0}
\end{equation}

$$\frac{d\sigma (e^+e^-\to
f_0\gamma\to\pi^0\pi^0\gamma)}{dm}=\frac{g_{f_0\pi^0\pi^0}^2
p_{\pi\pi}(m)}{4\pi^2|D_{f_0}(m)|^2}\,\sigma (e^+e^-\to
f_0\gamma,m)=$$ \begin{equation} \frac{e^2
(s-m^2)p_{\pi\pi}(m)}{96\pi^3s^3}|A_{K^+K^-}(s) \bar g(m)|^2\bigg|
\frac{g_{f_0
K^+K^-}g_{f_0\pi^0\pi^0}}{D_{f_0}(m)}\bigg|^2\,,\label{sigmaSignalf0}\end{equation}

$$p_{\eta\pi}(m)=\frac{\sqrt{(m^2-(m_\eta+m_\pi)^2)(m^2-(m_\eta-m_\pi)^2)}}{2m},\
\  p_{\pi\pi}(m)=\frac{\sqrt{m^2-4m_\pi^2}}{2}$$

The $A_{K^+K^-}(s)$ is the amplitude of the $\gamma^*\to K^+K^-$
transition. Without mixing of the intermediate vector states this
amplitude is

\begin{equation} A_{K^+K^-}(s)=\sum_{V=\rho \!,\,\rho '\!,\,\rho ''\!,\,
\omega\!,\,\omega '\!,\,\omega ''\!,\, \phi\!,\,\phi' \!,\,\phi''}
\frac{g_{V\gamma}g_{VK^+K^-}}{D_{V}(s)}\,, \end{equation}

We took into account mixing of the resonances according to Ref.
\cite{achkozh-97}. The $m$ is the $\eta\pi^0$ or $\pi^0\pi^0$
invariant mass correspondingly,
$g_{f_0\pi^0\pi^0}=g_{f_0\pi^+\pi^-}/\sqrt{2}$ for the pions
identity and $\bar g(m)=\frac{g_R(m)}{g_{\phi K^+K^-}g_{R
K^+K^-}}$ is the integral on the kaon loop \cite{achasov-89}. The
$D_{a_0}(m)$ and $D_{f_0}(m)$ are the inverse propagators of
scalar mesons, they are taken from \cite{our_a0} and
\cite{our_f0_2011} (2012 paper) correspondingly. All parameters of
the $a_0(980)$ and $f_0(980)$ are taken from \cite{our_a0} (Fit 1)
and \cite{our_f0_2011} (Fit 1 of the 2012 paper) correspondingly.
The Eqs. (\ref{sigmaSignala0}) and (\ref{sigmaSignalf0}) may be
rewritten as

$$\frac{d\sigma (e^+e^-\to
a_0\gamma\to\eta\pi^0\gamma)}{dm}=\frac{(s-m^2)p_{\eta\pi}(m)}{2\pi^2
\sqrt{s}(s-4m_{K^+}^2)^{3/2}} |\bar g(m)|^2\bigg|\frac{g_{a_0
K^+K^-}g_{a_0 \eta\pi^0}}{D_{a_0}(m)}\bigg|^2\times $$
\begin{equation} \sigma (e^+e^-\to K^+K^-)
\,,\label{sigmaFromKKa0}\end{equation}

$$\frac{d\sigma (e^+e^-\to
f_0\gamma\to\pi^0\pi^0\gamma)}{dm}=\frac{(s-m^2)p_{\pi\pi}(m)}{2\pi^2
\sqrt{s}(s-4m_{K^+}^2)^{3/2}} |\bar g(m)|^2 \bigg|\frac{g_{f_0
K^+K^-}g_{f_0\pi^0\pi^0}}{D_{f_0}(m)}\bigg|^2\times $$
\begin{equation} \sigma (e^+e^-\to
K^+K^-)\label{sigmaFromKKf0}\end{equation}

\begin{figure}
\begin{center}
\begin{tabular}{ccc}
\includegraphics[width=8cm]{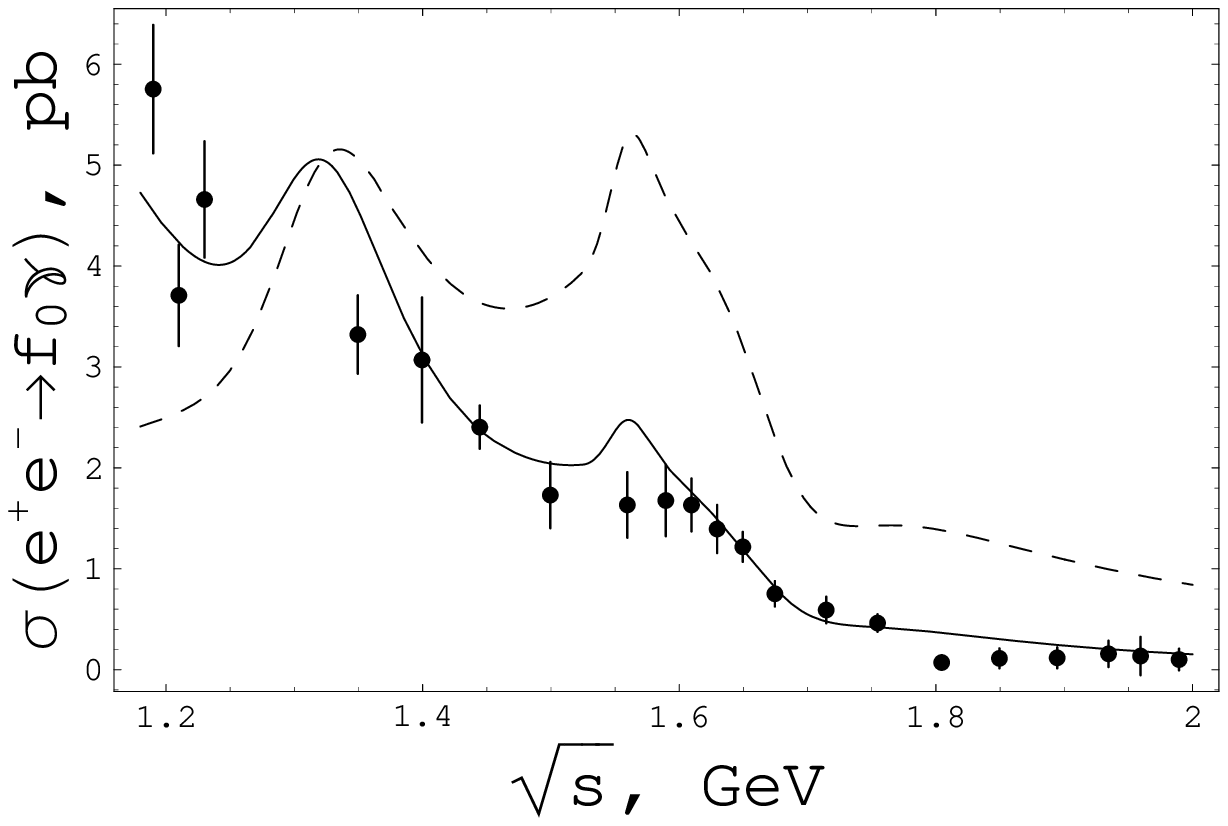}& \raisebox{-1mm}{$\includegraphics[width=8cm]{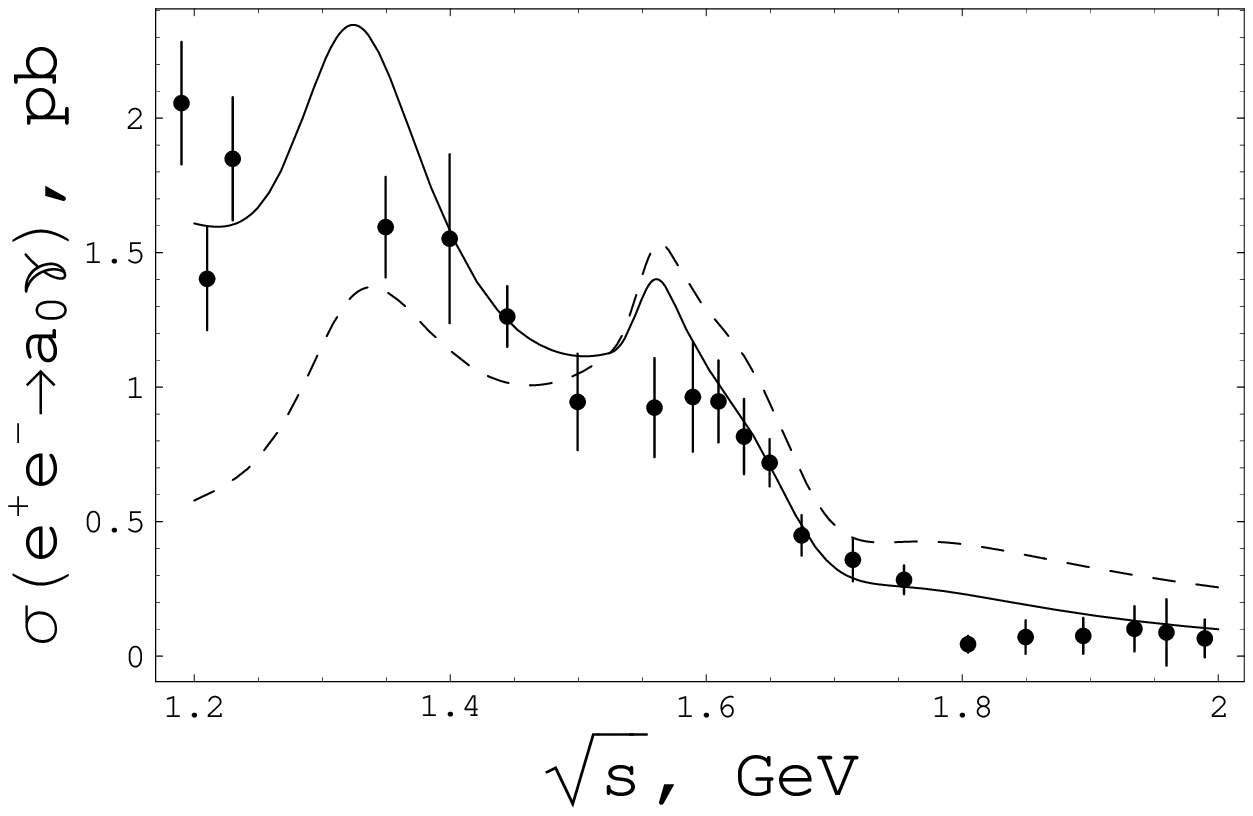}$}\\ (a)&(b)
\end{tabular}
\end{center}
\caption{The a) $\sigma (e^+e^-\to f_0\gamma )$ and b) $\sigma
(e^+e^-\to a_0\gamma )$. Solid lines show the kaon loop model
prediction, dashed lines show 1/10 of the cross-sections in case
of the mere VDM prediction for the point-like $VS\gamma$
interaction, normalized on the signal at $\sqrt{s}=m_\phi$. Points
are result of using the experimental data \cite{olyaKK,dm2KK} on
the $e^+e^-\to K^+K^-$ in the Eqs. (\ref{sigmaFromKKa0}),
(\ref{sigmaFromKKf0}).} \label{fig1}
\end{figure}

The photon angular distribution is

\begin{equation} \frac{dn^S_\gamma}{d\Omega}=\frac{3}{16\pi}
(1+\mbox{cos}^2\theta )\,, \end{equation}

\noindent where $\theta$ is the angle between the $\gamma$
momentum and the beam axis.

In Fig. \ref{fig1} we show the $\sigma (e^+e^-\to K^+K^-\to
S\gamma )$ cross-sections prediction, the mere VDM one (point-like
$VS\gamma$ interaction) is shown also \cite{mereVDM}. Points on
Fig. \ref{fig1} are obtained with the help of the experimental
data \cite{olyaKK,dm2KK} on the $e^+e^-\to K^+K^-$ and Eqs.
(\ref{sigmaFromKKa0}), (\ref{sigmaFromKKf0}). Note that $\sigma
(e^+e^-\to a_0\gamma \to \eta\pi^0\gamma)$ and $\sigma (e^+e^-\to
f_0\gamma \to \pi^0\pi^0\gamma)$ are considerably less for the
branching fractions and the energy cut of the narrow $a_0$ and
$f_0$ resonance regions.

Note also that the Vector Dominance Model describes formfactors (and
transition formfactors) of $q\bar q$ states in the low and
intermediate energy regions. In case of the four quark states
S=$a_0,f_0$ the amplitude of the process $\gamma^*\to S\gamma$
along with VDM suppression $(m_V^2-s)^{-1}$ has also additional
suppression $\frac{(2m_K)^2}{s}\ln^2\frac{s}{m_K^2}$ with
increasing $s$ for the kaon loop \cite{achasov-89prep}. This
provides additional suppression in comparison with the $q\bar q$ state case, see Fig. \ref{fig1}.

\section{The reaction $e^+e^-\to a_2 (f_2)\gamma$}

It is known that in the reaction $\gamma\gamma \to f_2\to
\pi\pi$ tenzor mesons are produced mainly by the photons with the
opposite helicity states. The effective Lagrangian in this case is

\begin{equation}
L=g_{f_2 \gamma\gamma}T_{\mu\nu}F_{\mu\sigma}F_{\nu\sigma}\,,\label{lagrfgg}
\end{equation}
$$F_{\mu\sigma}=\partial_\mu A_\sigma - \partial_\sigma A_\mu $$

\noindent where $A_{\mu}$ is a photon field and $T_{\mu\nu}$ is a
tenzor $f_2$ field. So in the frame of Vector Dominance Model
we assume that the effective Lagrangian of the reaction $f_2\to
VV$ is \cite{akar-86}:

\begin{equation}
L=g_{f_2 VV}T_{\mu\nu}F^V_{\mu\sigma}F^V_{\nu\sigma}\,,\label{lagrfvv}
\end{equation}
$$F^V_{\mu\sigma}=\partial_\mu V_\sigma-\partial_\sigma V_\mu \,
.$$

\noindent where
$V=\rho,\,\rho',\rho''\,\omega,\,\omega',\omega''$. Note that
$\omega,\,\omega'$ and $\omega''$ give $\sim 10\%$ of the
$\rho,\,\rho'$ and $\rho''$ contribution in the amplitude, so then
we neglect them: our current aim is to obtain estimates only. The
$\rho-\rho'-\rho''$ mixing is omitted here also.

Assuming the VDM mechanism $e^+e^-\to (\rho+\rho'+\rho'')\to f_2
(\rho+\rho'+\rho'')\to f_2 \gamma$ one obtains

$$\sigma(e^+e^-\to f_2\gamma)=\frac{4\pi^2}{9}\,\alpha^3
\Big(1-\frac{m_{f_2}^2}{s}\Big)^3\Big(\frac{s^2}{m_{f_2}^4}+3\frac{s}{m_{f_2}^2}+6\Big)\times
$$ \begin{equation} \bigg|\frac{m_\rho^2 g_{f_2\rho\rho}}{f^2_\rho
D_\rho(s)}+ \frac{m_{\rho'}^2 g_{f_2\rho'\rho'}}{f^2_{\rho'}
D_{\rho'}(s)}+ \frac{m_{\rho''}^2
g_{f_2\rho''\rho''}}{f^2_{\rho''} D_{\rho''}(s)}\bigg|^2
\label{crossf2} \end{equation}

\noindent and

\begin{equation}
\Gamma (f_2\to\gamma\gamma) =
\frac{\pi\alpha^2}{5}\bigg|\frac{g_{f_2\rho\rho}}{f^2_\rho}+
\frac{g_{f_2 \rho'\rho'}}{f^2_{\rho'}}+\frac{g_{f_2
\rho''\rho''}}{f^2_{\rho''}}\bigg|^2 m_{f_2}^3 = 3.03\pm 0.35\
\mbox{keV \cite{pdg-2012}}.
\end{equation}

We use the same parameters of the $\rho, \rho',\rho''$ as for the
$e^+e^-\to a_0\gamma$ and $e^+e^-\to f_0\gamma$ reactions. It is
assumed that $g_{f_2 \rho\rho'}$ and other cross constants are
suppressed due to small overlap of the spatial wave functions of
$\rho,\rho'$ and $\rho''$. The $f_V$ is obtained from the relation

\begin{equation}
\Gamma(V\to e^+e^-,s)=\frac{4\pi\alpha^2}{3f_V^2}\frac{m_V^4}{s^{3/2}}
\end{equation}

\begin{figure}
\begin{center}
\begin{tabular}{ccc}
\includegraphics[width=8cm]{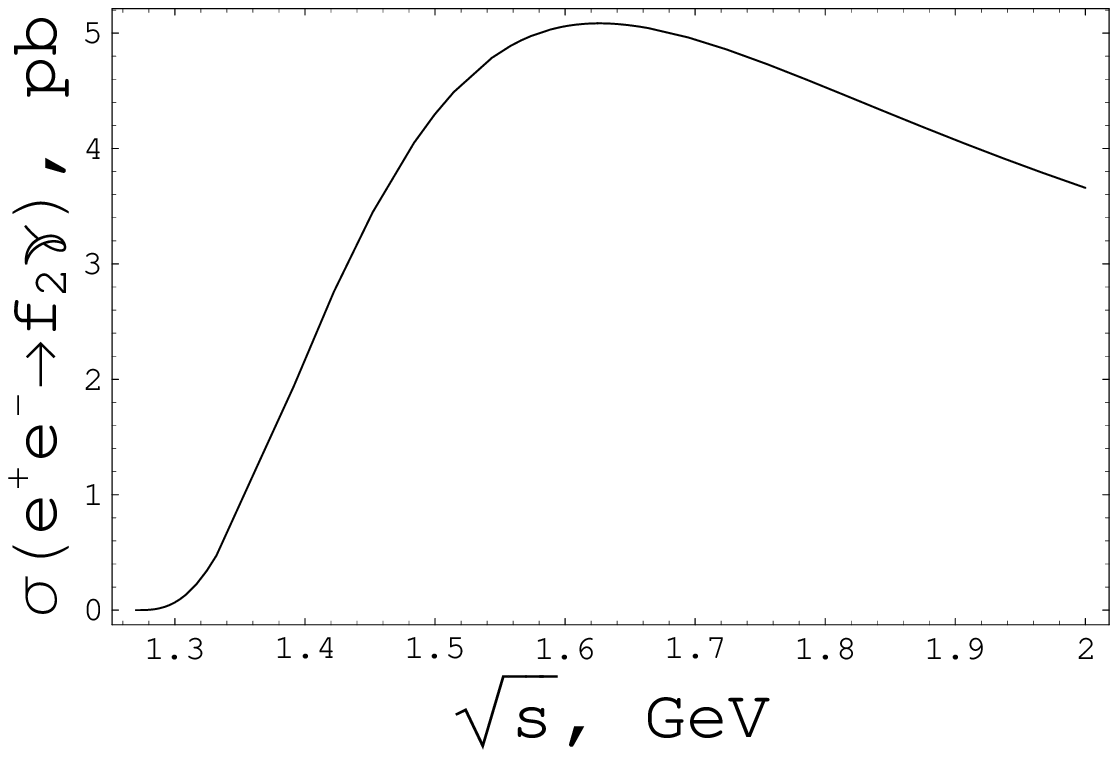}& \raisebox{-1mm}{$\includegraphics[width=8cm]{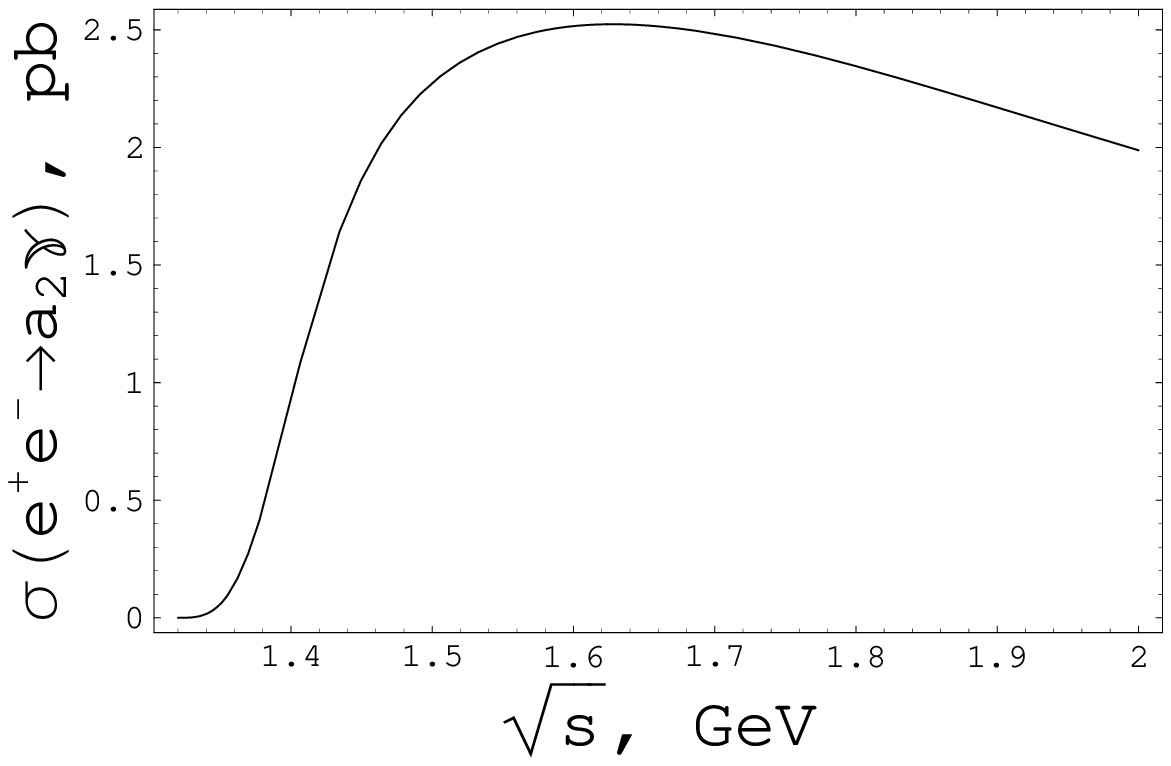}$}\\ (a)&(b)
\end{tabular}
\end{center}
\caption{a) The $\sigma (e^+e^-\to f_2\gamma)$ for
$g_{f_2\rho''\rho''}= g_{f_2\rho'\rho'}= g_{f_2\rho\rho}\,$. b)
The $\sigma (e^+e^-\to a_2\gamma)$ for
$g_{a_2\rho''\omega''}=g_{a_2\rho'\omega'}= g_{a_2\rho\omega}$.}
\label{figa2f2}
\end{figure}

If one assumes that $g_{f_2\rho\rho}=
g_{f_2\rho'\rho'}=g_{f_2\rho''\rho''}$, then the Fig.
\ref{figa2f2}a is obtained.

The $\pi^0\pi^0$ invariant mass $m$ spectra is
$$\frac{d\sigma(e^+e^-\to
f_2\gamma\to\pi^0\pi^0\gamma)}{dm}=\frac{8\pi}{9}\,\alpha^3
\Big(1-\frac{m^2}{s}\Big)^3\Big(\frac{s^2}{m_{f_2}^4}+3\frac{s}{m_{f_2}^2}+6\Big)\times
$$
\begin{equation}
\bigg|\frac{m_\rho^2 g_{f_2\rho\rho}}{f^2_\rho D_\rho(s)}+
\frac{m_{\rho'}^2 g_{f_2\rho'\rho'}}{f^2_{\rho'} D_{\rho'}(s)}+
\frac{m_{\rho''}^2 g_{f_2\rho''\rho''}}{f^2_{\rho''}
D_{\rho''}(s)}\bigg|^2\frac{\Gamma(f_2\to\pi^0\pi^0,m)m^2}{|D_{f_2}(m)|^2}
\end{equation}

\begin{figure}
\begin{center}
\begin{tabular}{ccc}
\includegraphics[width=8cm]{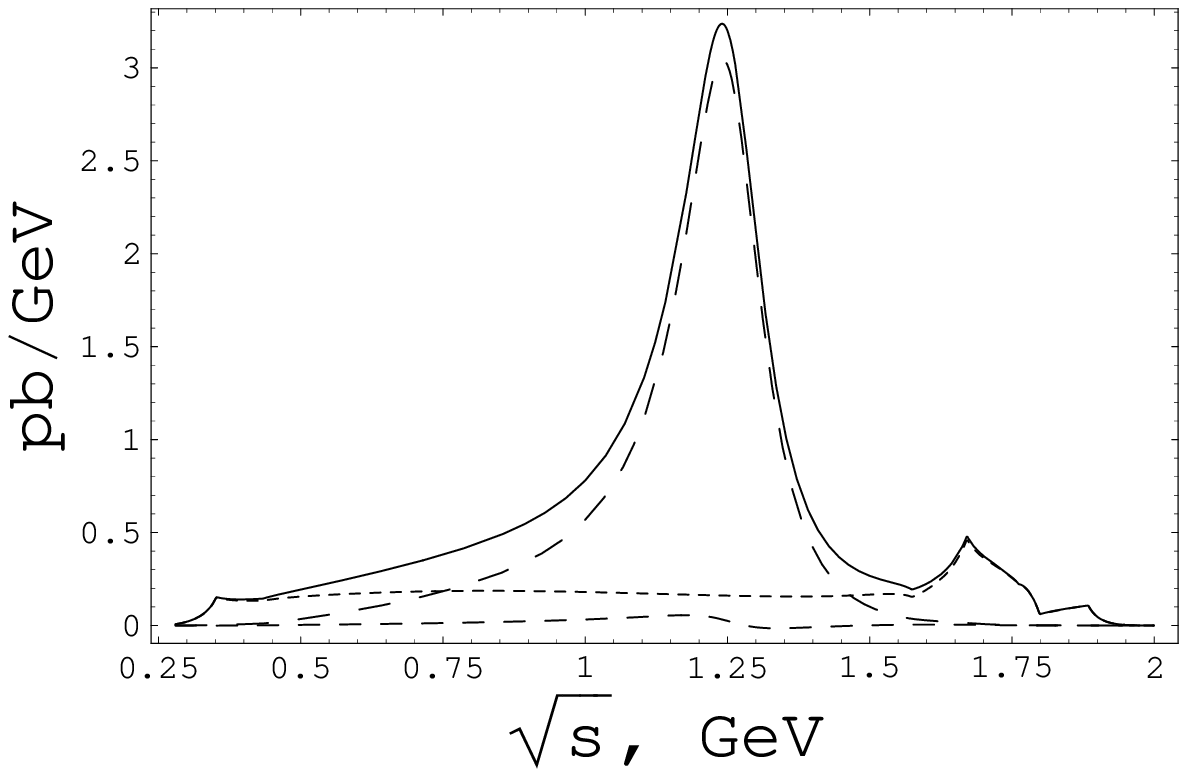}& \raisebox{-1mm}{$\includegraphics[width=8cm]{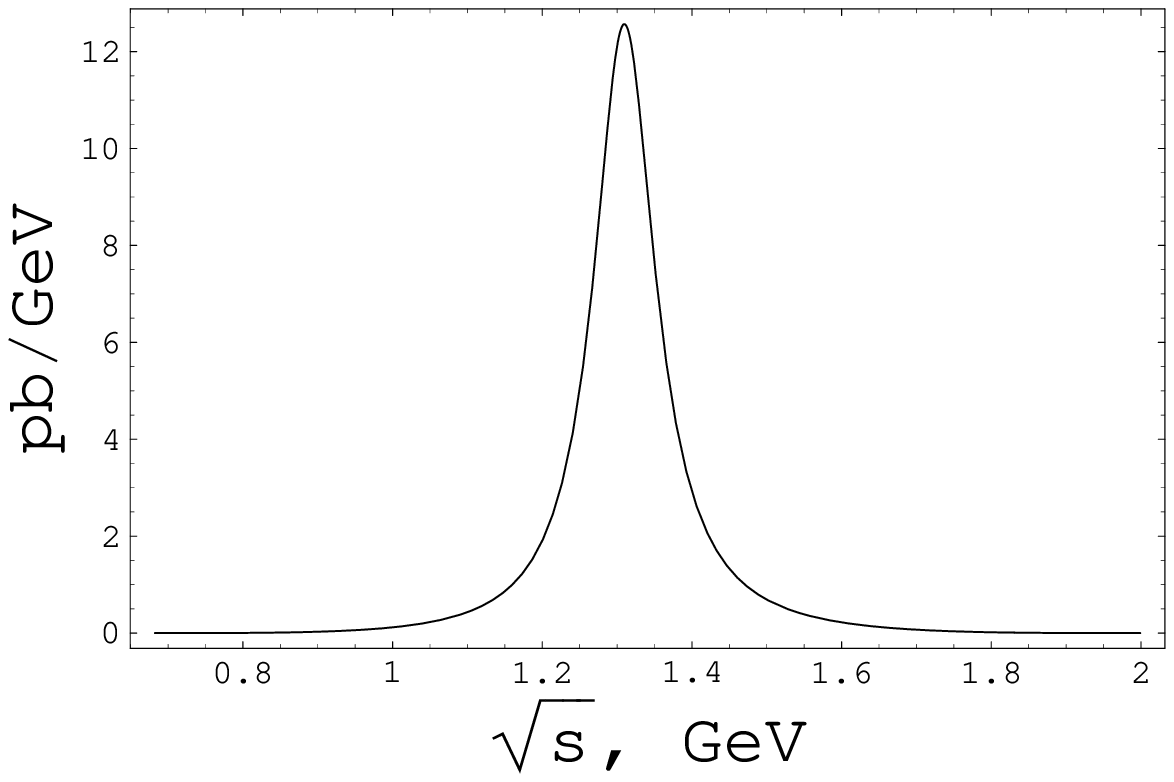}$}\\ (a)&(b)
\end{tabular}
\end{center}
\caption{a) The cutted $\pi^0\pi^0$ invariant mass spectra in the
reaction $e^+e^-\to\pi^0\pi^0\gamma$ for
$g_{f_2\rho''\rho''}=g_{f_2\rho'\rho'}=g_{f_2\rho\rho}$ at
$\sqrt{s}=2$ GeV. The cutting is 100 MeV around the $\omega$ mass
in $\pi^0(1)\gamma$ and $\pi^0(2)\gamma$ invariant masses. The
solid line is the sum of the cutted signal (long-dashed line),
cutted $\omega\pi^0$ background (short-dashed line) and the
interference (dashed line). b) The $\eta\pi^0$ invariant mass
spectra in the reaction $e^+e^-\to\eta\pi^0\gamma$ at $\sqrt{s}=2$
GeV for
$g_{a_2\omega''\rho''}=g_{a_2\omega'\rho'}=g_{a_2\omega\rho}$.}
\label{figSpec}
\end{figure}

Using the $\Gamma(f_2\to\pi^0\pi^0,m)$ and the inverse propagator
$D_{f_2}(m)$ from the Appendix, one gets Fig. \ref{figSpec}a,
where cutting $\pm 100$ MeV around $m_\omega$ in the $\pi^0\gamma$
mass was done to reduce the $\omega\pi^0$ background, see also
Sec. \ref{backSect}.

The photon angular distribution is

\begin{equation}
\frac{dn^T_\gamma}{d\Omega}= \frac{T}{N}
\end{equation}
\begin{equation}
T=\frac{(6m_f^4+s^2)(1+\mbox{cos}^2\theta)+6s\, m_f^2
\mbox{sin}^2\theta}{3m_f^4} \label{eqT}
\end{equation}
\begin{equation}
N=\frac{16\pi}{9}\Big(\frac{s^2}{m_{f_2}^4}+3\frac{s}{m_{f_2}^2}+6\Big)
\label{eqN}
\end{equation}

\begin{figure}
\begin{center}
\begin{tabular}{ccc}
\includegraphics[width=8cm]{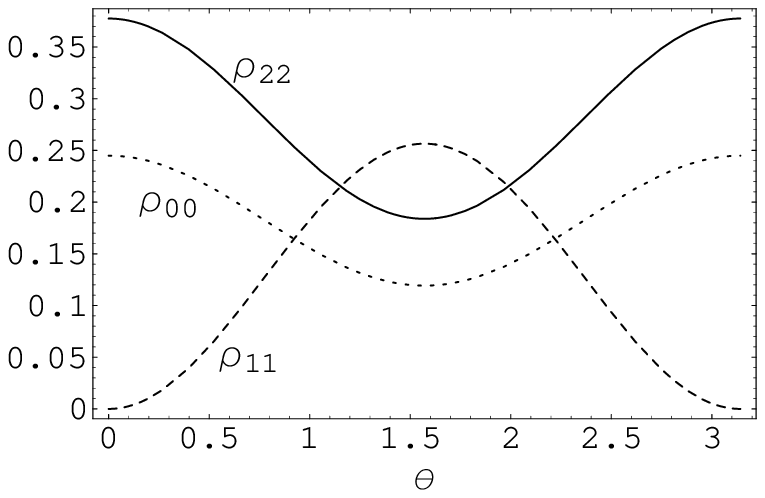}& \raisebox{-1mm}{$\includegraphics[width=8cm]{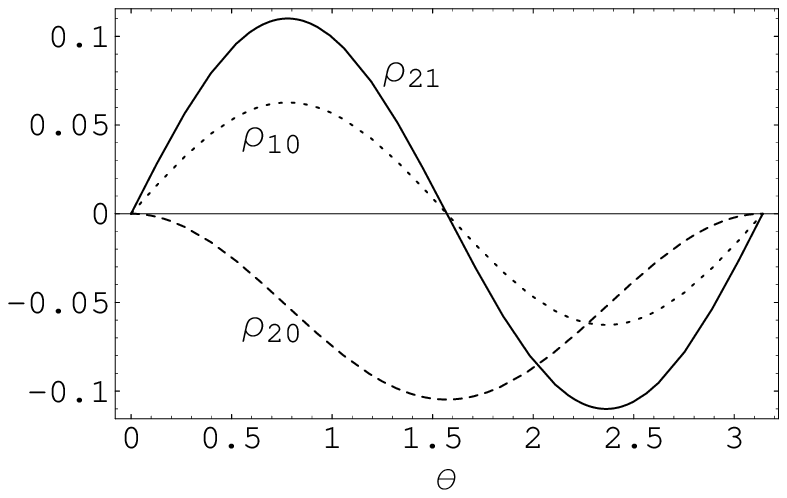}$}\\ (a)&(b)
\end{tabular}
\end{center}
\caption{The elements of the $f_2$ spin density matrix at $s=(1.5$
GeV$)^2$(see Table I): a) $\rho_{22}$ (solid line), $\rho_{11}$
(dashed line), $\rho_{00}$ (short-dashed line); b) $\rho_{21}$
(solid line), $\rho_{20}$ (dashed line), $\rho_{10}$ (short-dashed
line). } \label{fig7}
\end{figure}

\begin{figure}
\begin{center}
\begin{tabular}{ccc}
\includegraphics[width=8cm]{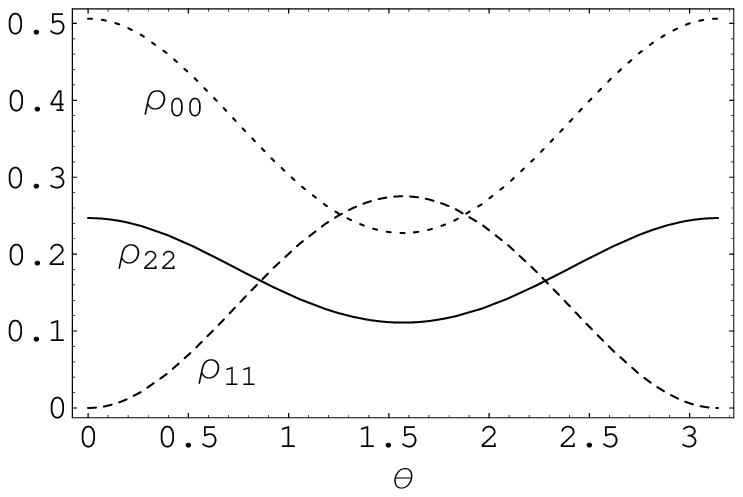}& \raisebox{-1mm}{$\includegraphics[width=8cm]{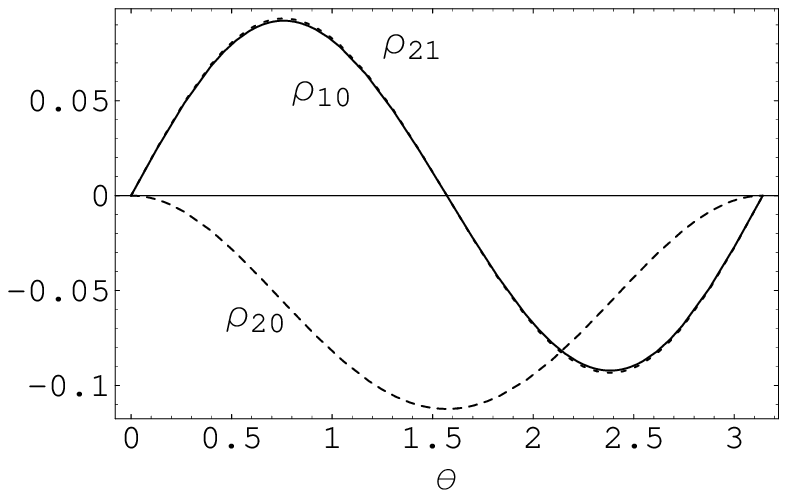}$}\\ (a)&(b)
\end{tabular}
\end{center}
\caption{The elements of the $f_2$ spin density matrix at $s=(2$
GeV$)^2$(see Table I): a) $\rho_{22}$ (solid line), $\rho_{11}$
(dashed line), $\rho_{00}$ (short-dashed line); b) $\rho_{21}$
(solid line), $\rho_{20}$ (dashed line), $\rho_{10}$ (short-dashed
line). } \label{fig8}
\end{figure}

\begin{center}
Table I. Elements of the $f_2$ spin density matrix, see Figs.
(\ref{fig7},\ref{fig8}) \vspace{5pt}

\begin{tabular}{|c|c|c|c|c|c|}\hline

$\rho_{22}$ & $\frac{1}{T}(1+\mbox{cos}^2\theta)$ & $\rho_{21}$ &
$\frac{1}{T}\frac{\sqrt{s}\mbox{sin}(2\theta)}{2m_f}$ &
$\rho_{20}$ &
$-\frac{1}{T}\frac{s\mbox{sin}^2\theta}{\sqrt{6}m_f^2}$
\\ \hline

$\rho_{11}$ & $\frac{1}{T}\frac{s\mbox{sin}^2\theta}{m_f^2}$ &
$\rho_{10}$ &
$\frac{1}{T}\frac{s^{3/2}\mbox{sin}(2\theta)}{2\sqrt{6}m_f^3}$ &
$\rho_{1,-1}$ & 0\\ \hline

$\rho_{00}$ &
$\frac{1}{T}\frac{s^2(1+\mbox{cos}^2\theta)}{3m_f^4}$ &
$\rho_{2,-1}$ & 0 & $\rho_{2,-2}$ & 0
\\ \hline

\end{tabular}
\end{center}

The pion angular distribution in the rest frame of the tensor
meson is
$$W_2(\vartheta,\varphi)=\frac{15}{16\pi}\bigg\{\mbox{sin}^4\vartheta\,\rho_{22}+\mbox{sin}^2
2\vartheta\,\rho_{11} + 3\bigg(\mbox{cos}^2
\vartheta-\frac{1}{3}\bigg)^2\,\rho_{00}$$
\begin{equation}
+2\mbox{cos}\varphi\,\mbox{sin}2\vartheta\,
\bigg(\mbox{sin}^2\vartheta\,\rho_{21} -\sqrt{6}\bigg(\mbox{cos}^2
\vartheta-\frac{1}{3}\bigg)\,\rho_{10}\bigg)-2\sqrt{6}\,\mbox{cos}2\varphi\,\mbox{sin}^2\vartheta\,
\bigg(\mbox{cos}^2 \vartheta-\frac{1}{3}\bigg)\,\rho_{20} \bigg\}
\label{angleDistr1}
\end{equation}

Here z axis is along the direction of the tenzor meson momentum in
the $e^+e^-$ s.c.m., x is in the reaction plane and y is
perpendicular to the reaction plane
\cite{gottfriedJackson,jackson}, $\vartheta$ is the polar angle
and $\varphi$ is the azimuthal angle.

After integration over $\varphi$ one has

\begin{equation}
W_2(\vartheta)=\frac{15}{8}\bigg(\mbox{sin}^4\vartheta\,\rho_{22}+\mbox{sin}^2
2\vartheta\,\rho_{11} + 3(\mbox{cos}^2
\vartheta-\frac{1}{3})^2\,\rho_{00}\bigg) \label{angleDistr2}
\end{equation}

For the $e^+e^-\to a_2\gamma$ we may use Lagrangians similar to
Eqs. (\ref{lagrfgg}), (\ref{lagrfvv}):

\begin{equation}
L=g_{a_2 \gamma\gamma}T_{\mu\nu}F_{\mu\sigma}F_{\nu\sigma}\,,\label{lagra2gg}
\end{equation}
\noindent and
\begin{equation}
L=g_{a_2
VV'}T_{\mu\nu}F^V_{\mu\sigma}F^{V'}_{\nu\sigma}\,,\label{lagra2vv}
\end{equation}

\noindent where $T_{\mu\nu}$ is the $a_2$ field,
$V=\rho,\rho',\rho''$ and $V'=\omega,\omega',\omega''$.

Analogically one may write $e^+e^-\to
(\rho+\omega+\rho'+\omega'+\rho''+\omega'')\to a_2
(\omega+\rho+\omega'+\rho'+\omega''+\rho'')\to a_2 \gamma$ cross
section, neglecting $a_2\rho\omega'$, $a_2\rho'\omega$ and other
cross vertices due to small overlap of the spatial wave functions
of $\rho,\,\omega'$ and $\rho',\,\omega$ and so on:

$$\sigma(e^+e^-\to a_2\gamma)=\frac{\pi^2}{9}\,\alpha^3
\Big(1-\frac{m_{a_2}^2}{s}\Big)^3\Big(\frac{s^2}{m_{a_2}^4}+3\frac{s}{m_{a_2}^2}+6\Big)\times
$$ \begin{equation}\bigg|\frac{m_\rho^2 g_{a_2\rho\omega}}{f_\rho
f_\omega D_\rho(s)}+ \frac{m_\omega^2 g_{a_2\rho\omega}}{f_\rho
f_\omega D_\omega(s)}+\frac{m_{\rho'}^2
g_{a_2{\rho'\omega'}}}{f_{\rho'} f_{\omega'} D_{\rho'}(s)}+
\frac{m_{\omega'}^2 g_{a_2\rho'\omega'}}{f_{\rho'} f_{\omega'}
D_{\omega'}(s)}+\frac{m_{\rho''}^2
g_{a_2{\rho''\omega''}}}{f_{\rho''} f_{\omega''} D_{\rho''}(s)}+
\frac{m_{\omega''}^2 g_{a_2\rho''\omega''}}{f_{\rho''}
f_{\omega''} D_{\omega''}(s)}\bigg|^2\label{crossa2}\end{equation}

\begin{equation} \Gamma (a_2\to\gamma\gamma) =
\frac{\pi\alpha^2}{5}\bigg|\frac{g_{a_2\rho\omega}}{f_\rho
f_\omega}+\frac{g_{a_2\rho'\omega'}}{f_{\rho'}
f_{\omega'}}+\frac{g_{a_2\rho''\omega''}}{f_{\rho''}
f_{\omega''}}\bigg|^2 m_{a_2}^3 = 1.00\pm 0.06\ \mbox{keV
\cite{pdg-2012}}.\end{equation}

Assuming
$g_{a_2\rho''\omega''}=g_{a_2\rho'\omega'}=g_{a_2\rho\omega}$
gives Fig. \ref{figa2f2}b and $\eta\pi^0$ spectrum shown in Fig.
\ref{figSpec}b. The angular distributions are the same as Eqs.
(\ref{angleDistr1}) and (\ref{angleDistr2}) (with $m_{f_2}\to
m_{a_2}$ substitution in Eqs. (\ref{eqT}) and (\ref{eqN})).









\begin{figure}
\begin{center}
\begin{tabular}{ccc}
\includegraphics[width=10cm]{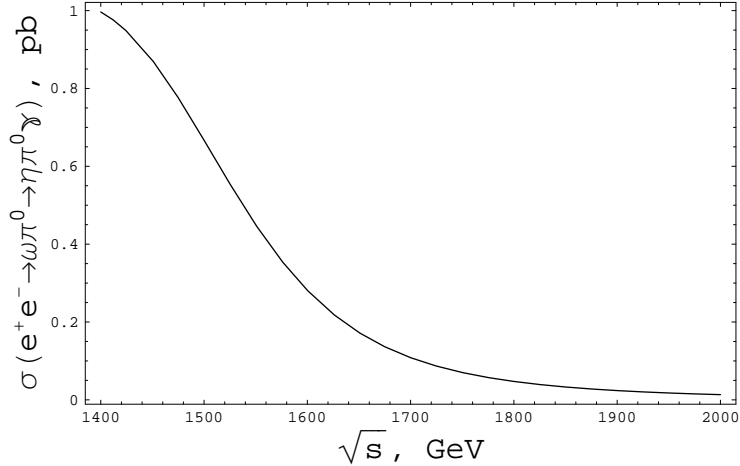}
\end{tabular}
\end{center}
\caption{The background cross-section
$\sigma(e^+e^-\to\omega\pi^0\to\eta\pi^0\gamma)$, obtained with
the help of Ref. \cite{achkozh-96}.} \label{backa2}
\end{figure}

\section{The background situation}\label{backSect}

Because of weakness of the signal cross-sections in the region
$1.4\div 2$ GeV it is needed to treat the background situation
accurately. The $e^+e^-\to f_0\gamma\to \pi^0\pi^0\gamma$ and
$e^+e^-\to a_0\gamma\to \eta\pi^0\gamma$ are expected to be too
small to be observed themselves, but for $e^+e^-\to f_2\gamma\to
\pi^0\pi^0\gamma$ and $e^+e^-\to a_2\gamma\to \eta\pi^0\gamma$ it
is possible to reduce background to the level much less than the
signal one.

The background situation in case of $e^+e^-\to
a_0\gamma\to\eta\pi^0\gamma$ and $e^+e^-\to a_2\gamma
\to\eta\pi^0\gamma$ is rather friendly. The main background
$e^+e^-\to\omega\pi^0\to\eta\pi^0\gamma$, obtained with the help
of Ref. \cite{achkozh-96}, is much less than the signal
cross-section, see Figs. \ref{figa2f2}b, \ref{backa2}. Other
backgrounds are even much smaller.

The main background in case of $e^+e^-\to
f_0\gamma\to\pi^0\pi^0\gamma$ and $e^+e^-\to f_2\gamma
\to\pi^0\pi^0\gamma$ reactions comes from the $\omega \pi^0$
intermediate state also. Fig. \ref{figSpec}a shows that cut $\pm
100$ MeV from the $\omega$ mass in the $\pi^0\gamma$ distribution
reduces this background to the small level.

\section{Conclusion}
Our analysis shows that it should be possible to observe the reactions
$e^+e^-\to a_2\gamma\to\eta\pi^0\gamma$ and $e^+e^-\to
f_2\gamma\to\pi^0\pi^0\gamma $ at the energies near $2$ GeV in
VEPP-2000 after reaching the project luminosity and probably in
$DA\Phi NE$ after the planned full upgrade.

As to the reactions $e^+e^-\to a_0\gamma\to\eta\pi^0\gamma$ and
$e^+e^-\to f_0\gamma\to\pi^0\pi^0\gamma$ at the energies $1.7-2$
GeV, in the kaon loop model their cross sections are very small,
while if the mere VDM (point-like $VS\gamma$ interaction) was
correct they would be observed, see Fig. \ref{fig1}. Consequently,
observation of tenzor contribution and non-observation of the
scalar one would support the four quark nature of light scalar
mesons \cite{jaffe1,jaffe2,z_phys}.

\section{Acknowledgements}

We thank A.A. Kozhevnikov very much for useful discussions
concerning $e^+e^-\to K^+K^-$ cross-section, drawn in Ref.
\cite{achkozh-97}.

This work was supported in part by RFBR, Grant No. 13-02-00039,
and Interdisciplinary Project No. 102 of the Siberian division of
RAS.

\section{Appendix: The propagators of the $a_2(1320)$ and $f_2(1270)$}
\label{Tprops}

According to \cite{annsgn2011}, we used the following inverse
propagators for $T=a_2,\,f_2$: \begin{equation} D_{T}(m^2)
=m_T^2-m^2-im\Gamma_T(m)\,,\end{equation} \noindent where

\begin{equation}\Gamma_{a_2}(m)=\Gamma(a_2\to\pi\pi,m)=
\Gamma_{a_2}^{tot}\frac{m_{a_2}^2}{m^2}\frac{p^5_{\eta\pi}(m)}
{p^5_{\eta\pi}(m_{a_2})}\frac{D_2(r_{a_2}p_{\eta\pi}(m_{a_2}))}{D_2(r_{a_2}p_{\eta\pi}(m))}\,,
\end{equation} \noindent and
\begin{equation} \Gamma_{f_2}(m)=\Gamma(f_2\to\pi\pi,m)+\Gamma(f_2\to K\bar
K,m)+\Gamma(f_2\to 4\pi,m)\,,\end{equation} \noindent which is
dominated by
\begin{equation}
\Gamma(f_2\to\pi\pi,m)=\Gamma_{f_2}^{tot}B(f_2\to\pi\pi)\frac{m_{f_2}^2}{m^2}\frac{p^5_{\pi\pi}(m)}{p^5_{\pi\pi}(m_{f_2})}\frac{D_2(r_{f_2}p_{\pi\pi}(m_{f_2}))}{D_2(r_{f_2}p_{\pi\pi}(m))}\,.
\label {f2PP}
\end{equation}

Here $D_2(x)=9+3x^2+x^4$ \cite{Blatt-Weisskopf}, and we take from
Ref. \cite{annsgn2011} $m_{a_2}=1322$ MeV,
$\Gamma_{a_2}^{tot}=116$ MeV, $r_{a_2}=1.9$ GeV$^{-1}$,
$m_{f_2}=1272$ MeV, $\Gamma_{f_2}^{tot}=196$ MeV,
$B(f_2\to\pi\pi)=0.848$, $r_{f_2}=8.2$ GeV$^{-1}$. The
$\Gamma(f_2\to K\bar K,m)$ has the similar form as Eq.
(\ref{f2PP}), we use $B(f_2\to K\bar K)=0.046$. The $\Gamma(f_2\to
4\pi,m)$ may be approximated by the S-wave $f_2\to\rho\rho\to
4\pi$ decay width as in Ref. \cite{annsgn2011}, but for simplicity
we used the dependence $\Gamma(f_2\to 4\pi,m)=\Gamma(f_2\to
4\pi,m_{f_2})\frac{m^2}{m_{f_2}^2}$ as in Ref. \cite{uehara}, here
$B(f_2\to 4\pi)=0.106$.


\begin{thebibliography}{99}
\bibitem{mereVDM}
The mere VDM prediction is obtained by the removement of the loop
integral $\bar g(m)$ and normalization of the cross-section on the
signal one Eqs. (\ref{sigmaSignala0}), (\ref{sigmaSignalf0}) at
$\sqrt{s}=m_\phi$.
\end{thebibliography}
\end{document}